# Effect of Subduction Zones on the Structure of the Small-Scale Currents at Core-Mantle Boundary.

Sergey Ivanov[1], Irina Demina[1], Sergey Merkuryev[1,2]

[1]St-Petersburg Filial of Pushkov Institute of Terrestrial Magnetism, Ionosphere and Radio wave Propagation (SPbF IZMIRAN)
[2]Saint Petersburg State University, Institute of Earth Sciences
Universitetskaya nab., 7-9,
St. Petersburg, 199034, Russia
`dim@izmiran.spb.ru`

**Abstract.** The purpose of this work is to compare kinematics of small-scale current vortices located near the core-mantle boundary with high-speed anomalies of seismic wave velocity in the lowest mantle associated with the subduction zones. The small-scale vortex paths were early obtained by the authors in the frame of the macro model of the main geomagnetic field sources. Two sources were chosen whose kinematics are characterized by the complete absence of the western drift and whose paths have a very complex shape. Both sources are located in the vicinity of the subduction zones characterized by the extensive coherent regions with increased speed of seismic waves in the lowest mantle. One of them is geographically located near the western coast of Canada and the second one is located in the vicinity of Sumatra. For this study we used the global models of the heterogeneities of seismic wave velocity. It was obtained that the complex trajectories of the vortices is fully consistent with the high-speed anomalies of seismic wave velocity in the lowest mantle. It can be assumed that mixing up with the matter of the lowest mantle, the substance of the liquid core rises along the lowest mantle channel and promotes its further increase. In addition, the volume of oceanic crust, subducted millions of years ago, turned out to be sufficient to penetrate into the liquid core, forming the complex shape restrictions for free circulation of the core liquid.

## 1    Introduction

Numerous studies of the internal structure of the Earth by seismic tomography have revealed the complex structure of both the upper and lower mantles. Seismic tomography is a technique that uses seismic waves from earthquakes to image Earth's internal structure [1, 2]. The mantle heterogeneity manifest themselves by laterally varying in seismic wave velocity relative to the mean value for a given layer. The higher and lower values of seismic S and P wave velocities are expressed in percentage of the average and used to build models. Based on the accumulated data, global models of Earth's structure are constructed for both the upper and lower mantles [3-8].However, both the actual structure of the mantle and the interpretation of its heterogeneities are presently widely debated.

Another approach concerning this issue is taken in this paper. We used the author's macro model of the main geomagnetic field's sources [9,10], in particular, the kinematics of individual small-scale current vortices localized near the core-mantle boundary. The analyzed vortices were obtained as a part of the macromodel of the main geomagnetic field sources. This model was built on the basis of the spatial structure of the main geomagnetic field components, calculated from the coefficients of the spherical harmonic IGRF model [11] for the period 1900-2010 and the gufm1 model [12]for earlier epochs. The problem was solved by the techniques of sequential separation of the contributions of different order sources, starting with the most powerful, followed by an iterative process to specify the parameters of previously selected sources and to select new ones. For each source, the parameters were determined by the least squares method, the minimum of the nonlinear functional was found by the Nelder-Mead algorithm. The problem was solved independently for each epoch, which allowed to construct the parameters time series, in particular, the vortex path. We understand by a vortex path the changes the position of the source relative to the Earth's surface. The vortex path projections on the Earth surface are shown in Fig.1 for the part of them.

As it can seen on Fig.1 the shape of the vortices paths are too complicated to be explained by the differential rotation of the different shells of the Earth only. Since differential rotation in the first plane would manifest itself as western drift of vortices. We have assumed that the shape of the vortex path is primarily determined by the heterogeneity of the structure of the lowest mantle and the features of the core-mantle boundary structure.

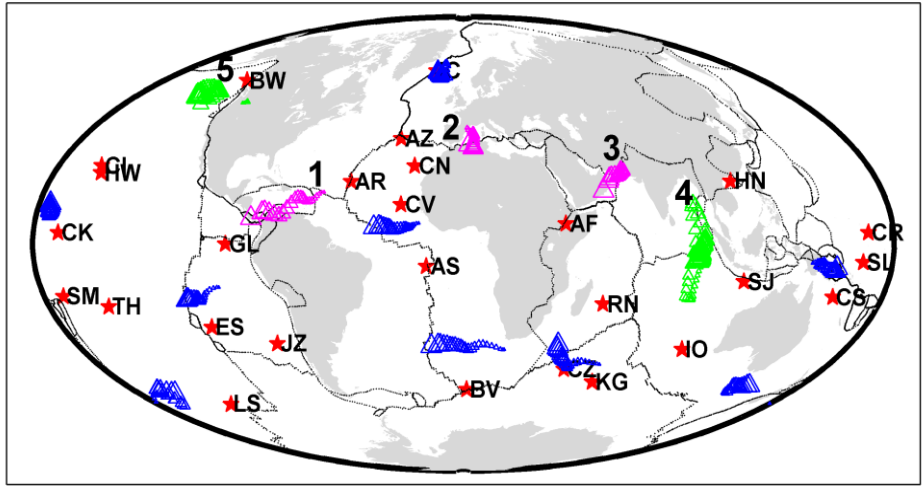

**Fig. 1.** Paths of small-scale vortices projected on the Earth's surface. Path points are shown by △ The symbol size is proportional to time. Boundaries of the lithospheric plates are shown by grey line, low mantel plumes are shown by ★, AR is the plume name.

We have already carried out similar studies for the Caribbean region [13] and for regions of two foci of the secular variation of the main geomagnetic field (European and Caspian regions [14]). In the first case the kinematics of the corresponding vortex marked with 1 and magenta color in Fig.1 in general has a pronounced western drift in spite of its individual features. In [13] it was obtained that all these features are related to the structure of high-speed seismic wave anomalies in the lowest mantle. Regarding the other two vortices marked with magenta color and 2, 3 respectively in Fig.1, its paths are characterized by the complete absence of the western drift. And the kinematics of the vortices is completely dependent on the structure of the lowest mantle.

In this paper we have chosen two small-scale symmetric current vortex, whose kinematics are characterized by the complete absence of the western drift, and whose paths have a very complex shape. One of them is geographically located near the western coast of Canada (Canadian vortex) and the second one is located in the vicinity of Sumatra (Sumatran vortex). The geographical positions of these vortex trajectories served as an additional criterion for their selection, since both areas belong to subduction zones and are characterized by the presence of the extensive connected regions with increased speed of seismic waves in the lowest mantle that indicates the presence of a denser and colder substance. Both these areas belong to the so-called "cemeteries" of the lithospheric plates [15].

The geography of the regions under consideration is shown in Fig. 2 together with the tectonic plate boundaries, subduction zones, and other tectonic features.

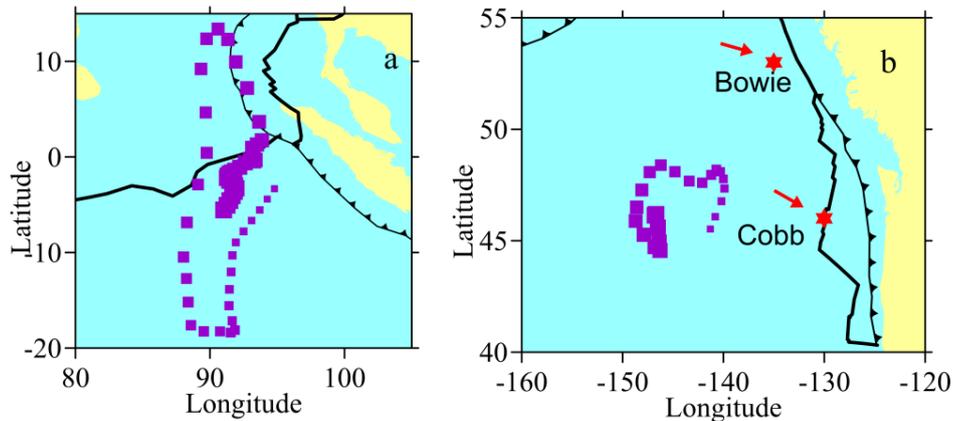

**Fig. 2.** Geographical location of the sources. a- Sumatran source; b- Canadian source;
■ projection of the source trajectory on the Earth surface; —— plates boundaries; ⊶ subduction zone; ★ hotspot

In this study we have selected the SAW642AN model for the calculation of the structure of the lowest mantle for several reasons. First, we already used the previous version of this model (SAW24) to analyze the Caspian and Europe-an foci of the main geomagnetic field [14]. Secondly, the SAW642AN model has the high resolution of in the lowest mantle. Besides, this is one of the latest global models, it is easily accessible for use and allows to produce slice-by-slice calculations at any depth with any incremental step.

## 2 Sumatran vortex

According to the tectonic type, the region around the island of Sumatra belongs to the island arcs. The subduction zone can occur at the oceanic-continental crust boundary. In this process, the oceanic crust always slides under the continental crust because of its greater density. The island arc is characterized by involving in subduction process of two different oceanic lithospheric plates, with one of them sliding under the other [16].

The downgoing plate can be traced in the upper mantle. At the same time, if the spreading rates were sufficiently high in the distant past, the remains of the subducted slabs can be detected in the lowest mantle in a form of regions with higher values of the seismic wave velocity.

Several slices of the distribution of the seismic wave velocity inhomogeneities, which were calculated using the SAW642AN model are shown in Fig. 3. The color scale displays the increased and lowered values of the seismic wave velocities.

An extensive high-speed anomaly is clearly distinguished on the slice (Fig. 3) corresponding to a depth of 2900 km which is assumed core-mantle boundary. The continuation of this anomaly can be traced higher as well. At the same time, several anomalies of lower speeds can be seen. These small size anomalies cut a high-speed anomaly, forming narrow lower-mantle channels.The Sumatra vortex trajectory also shown in Fig. 3. We slightly smoothed the curve of the trajectory for presentation on the fig. 3.

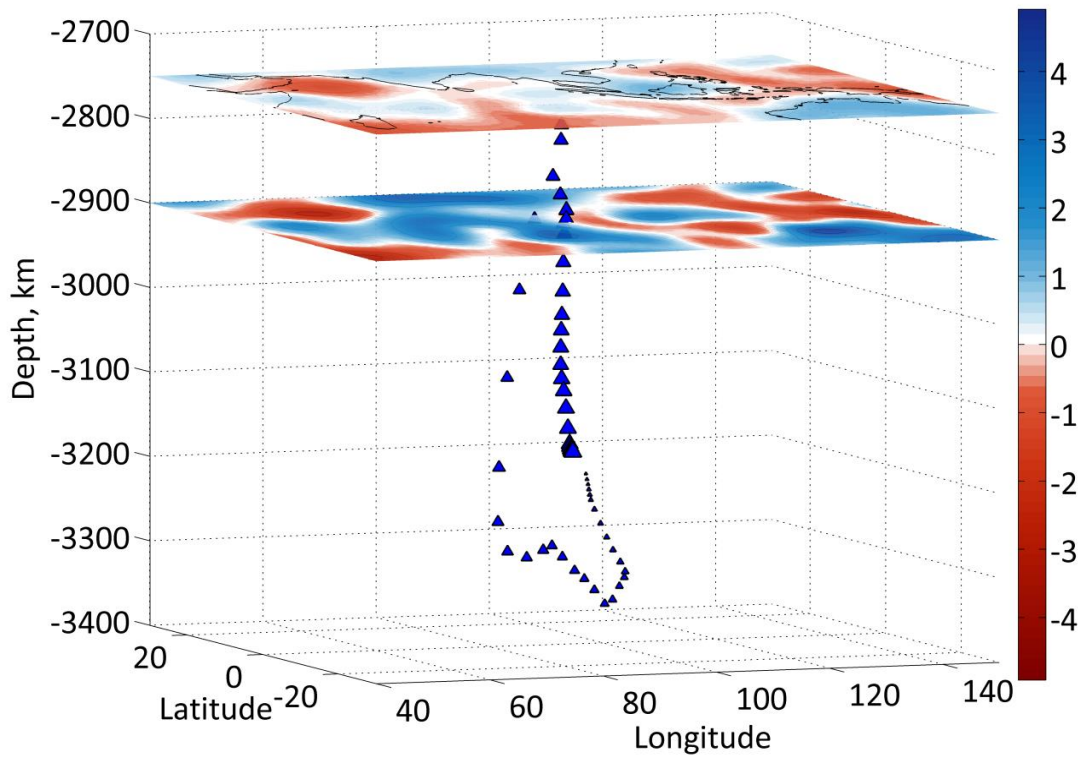

**Fig.3.** Color charts at depths 2750 and 2900 km are calculated using the SAW642AN model. In this and all the following figures the red and blue colors denote low-velocity and high-velocity anomalies, respectively; black line is the geographical outline, ▲ is the smoothed path of the vortex.

The vortex trajectory was obtained for the 400-year interval of 1600-2010 yrs. Almost throughout the entire time interval, with the exception of several points, the trajectory passes below the core-mantle boundary. However, the western drift is completely absent. Longitude remains virtually unchanged. The main movement is associated with a change in latitude. It can be seen that when the source location coincides with one of the lowest mantle channels, the source begins to rise. It can be assumed that the formation of the channel is due to the mixing of the mantle substance with the substance of the liquid core. As a result an upward flow is formed and the source rises with it.

After reaching a depth of about 2750 km, the vortex started to move downward. This clearly corresponds to channel attenuation, a decrease in temperature and an increase in density. However, even after this loop in the kinematics of the vortex there is no movement to the west, as if the "block" of the subducted slabs continued downward and created a wall. On the assumption that the amount of lithospheric matter has decreased by a factor of 1.5 to a depth of 3200 km. we can construct a hypothetical two slices at the depth of 3050 km and 3180 km. Then the shape of the vortex path gets a simple logical explanation. This hypothetical planes are added in Fig. 4.

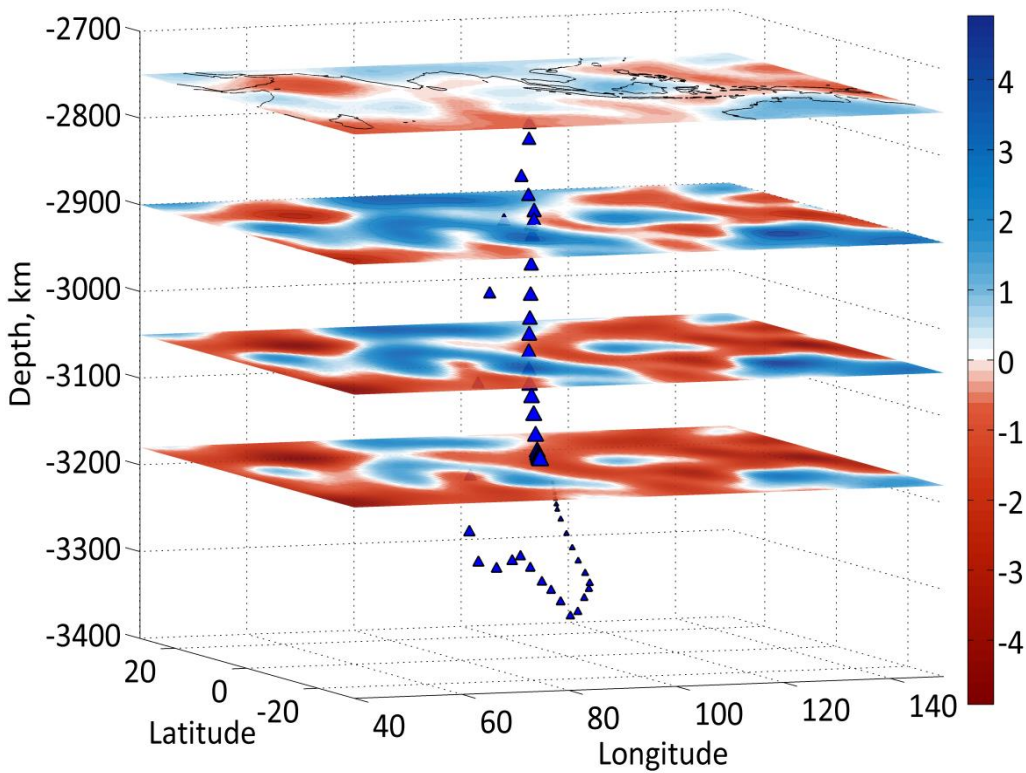

**Fig.4.** The hypothetical continuation of distribution of the subducted substance below the assumed core-mantle boundary at a depth of 3050 and 3200 km

At the same time,

Large volumes of residuals of subducted slabs were found near the core-mantle boundary by different authors. So in [17] such anomaly was detected as a result of detailed study of the Caribbean region. In [18], the authors note that these mass anomalies may not affect the shape of the geoid, but they affect the topography of the core-mantle boundary. Mountains were found on the border of 660 km, between the lower and upper mantle, by authors [19]. And in [20] the topography irregularities with a height of 14 km were found at the Earth's inner core boundary by seismic methods.

We are aware that our result is only a hypothesis. The accuracy of the path calculations is also limited by the knowledge of the main geomagnetic field in the past and the methodological difficulties in isolating individual sources. Therefore, we cannot definitely say that our numerical estimates accurately reflect the dimensions of the features of the topography of the core-mantle boundary. But the results suggest that the configuration of the core-mantle boundary is much more complicated than it is assumed to be at the present time. A similar result was drawn by us for Caribbean region. The Caribbean vortex path was obtained coinciding with the shape of the channel of lower-mantle plume. However, the most complex part of its morphology corresponds to the area below the core–mantle boundary, where it could be expected that the trajectory transits into the free westward drift. But if to suggest the possible penetration of the relics of the ancient tectonic plates into the liquid core in area below the Caribbean, then the vortex path could be explained.

## 3    Canadian vortex

The Canadian vortex region belongs to the zone of the San Andreas Fault (SAF). This area has a long tectonic history. Currently, SAF separates the North American and the Pacific plates. By its type SAF belongs to strick-slip faults, where the two fault sides shift predominantly horizontally relative to each other along the fault line, i.e. there is a simple off-set.

The SAF arose after the Farallon plate was subducted many millions of years ago. The Farallon plate was located west of the coast of North America and shared the modern sides of SAF. The relative motion of the Farallon and the North American plates was convergent. And the convergence speed was much higher than the spreading rate between the Pacific and the Farallon plates. As a result, the Farallon was completely subducted and the huge subducted masses of a lithosphere substance sank into the mantle. At present, a high-speed anomaly of seismic wave velocity has been found in the lowest mantle. This anomaly may be generated by the remnants of the Farallon plate.

In addition, this region is distinguished by two hotspots: Bowie and Cobb, which can be suggested as lower-mantle plumes. In Fig. 2 they are marked by red arrows. These hotspots are interesting in the framework of our problem, because, according to the existing assumptions, their roots go back to the core-mantle boundary, i.e. they belong to the lower-mantle plumes. Figure 5 shows several slices of the distribution of the seismic wave velocity inhomogeneities, calculated using the SAW642AN model.

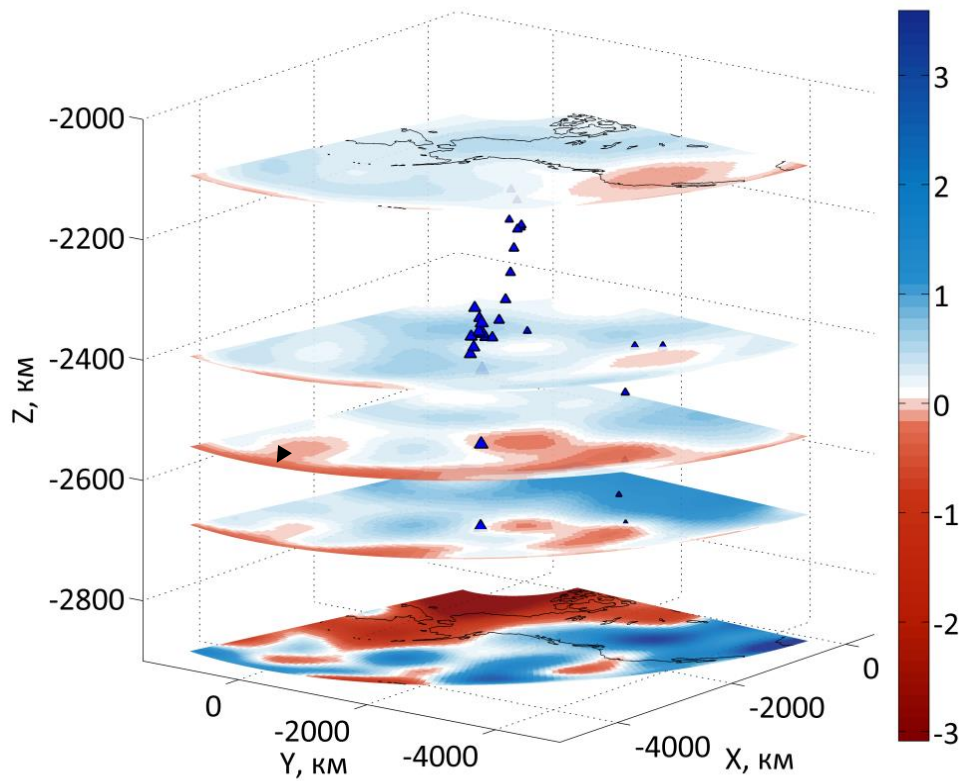

**Fig.5.** Color charts at depths of 2100, 2490, 2680 and 2890 km are calculated using the SAW642AN model. Legend is the same as for Fig. 3

The lowest slice corresponds to the core-mantle boundary. There is a powerful high-speed anomaly which is cut through by several thin low-speed channels at the depth of 2680 km. These channels can be roots of the hotspots. In our study, we do not solve the problem of tracing hot channels from the core-mantle boundary to the Earth surface. We are interested in how the source trajectory fits into this structure.

Unfortunately, we couldn't detect the Canadian vortex during the entire time period from year 1600 to 2010. Its path is shown in Fig. 5 with a large discontinuity. Because of the size of this gap, the unification of two segments into one path might raise questions, but this does not contradict the density structure of the lowest mantle. The first part of the path is an ascent along the lower-mantle channel corresponding to the Cobb hotspot. The last part passes along the lower-mantle channel corresponding to the Bowie hotspot. The configuration of the path between the depths of 2100 and 2500 km is very complicated. Direct comparison between the vortex path and the channel form is hampered by the fact that the SAW642AN model formally "loses" the low-speed Bowie channel on this depth interval. The configuration of the contour line of equal value remains in the speed inhomogeneity map, but part of them go to the positive range. To trace the channel configuration, you can additionally use other global models SMEAN2 and SAVANI [4], for example.

The SMEAN2 is a composite mantle tomography model composed SAVANI, GyPSuM [5], and S40RTS [6] generated using the approach of Becker and Boschi [7]. The slices calculated using the models SAVANI and SMEAN2 are shown in Fig. 6 and Fig. 7 in respective to the refined depth interval.

It is easy to see that both hot channels are well traced at these depths. The features of the source trajectory can be associated with a more complex configuration of the low-speed channel corresponding to the Bowie hotspot. This channel's shape is averaged in SAW642AN model at the depth between 2100 and 2500 km and cannot be distinctly traced. Thus, the kinematics of small-scale vortices located near the core-mantle boundary can be used as an additional indicator of the presence of low-speed ascending channels in the lowest mantle.

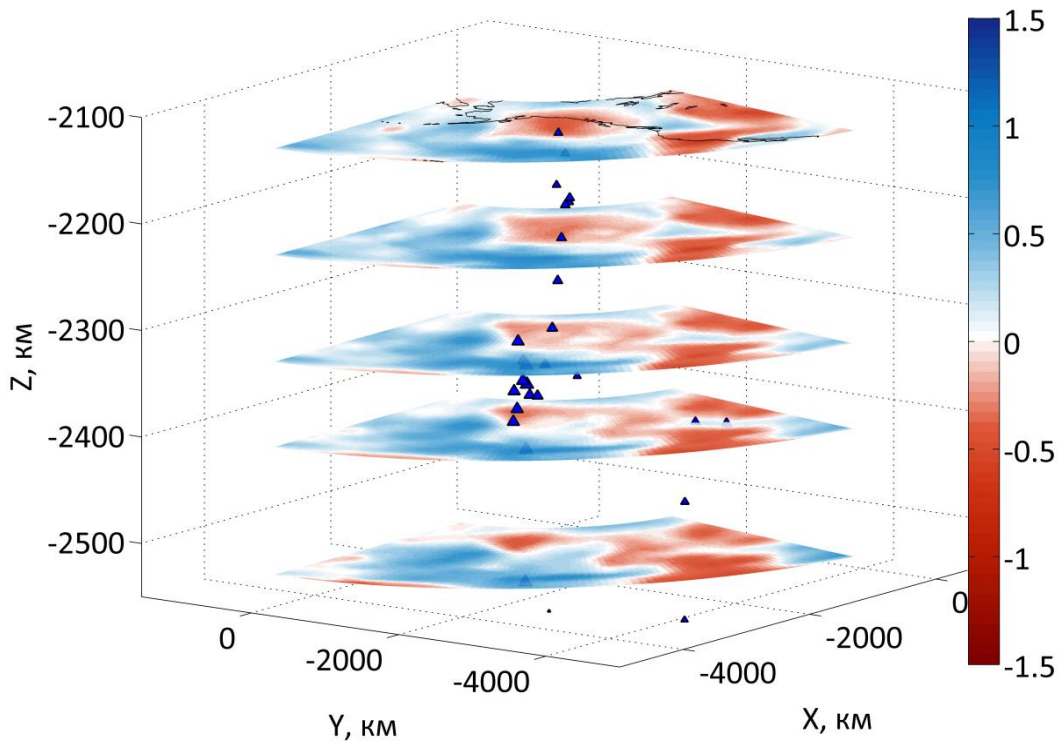

**Fig.6**. The lower-mantle structure calculated using the SAVANI model at the depth interval of 2150 to 2520 km. Legend is the same as for Fig. 3

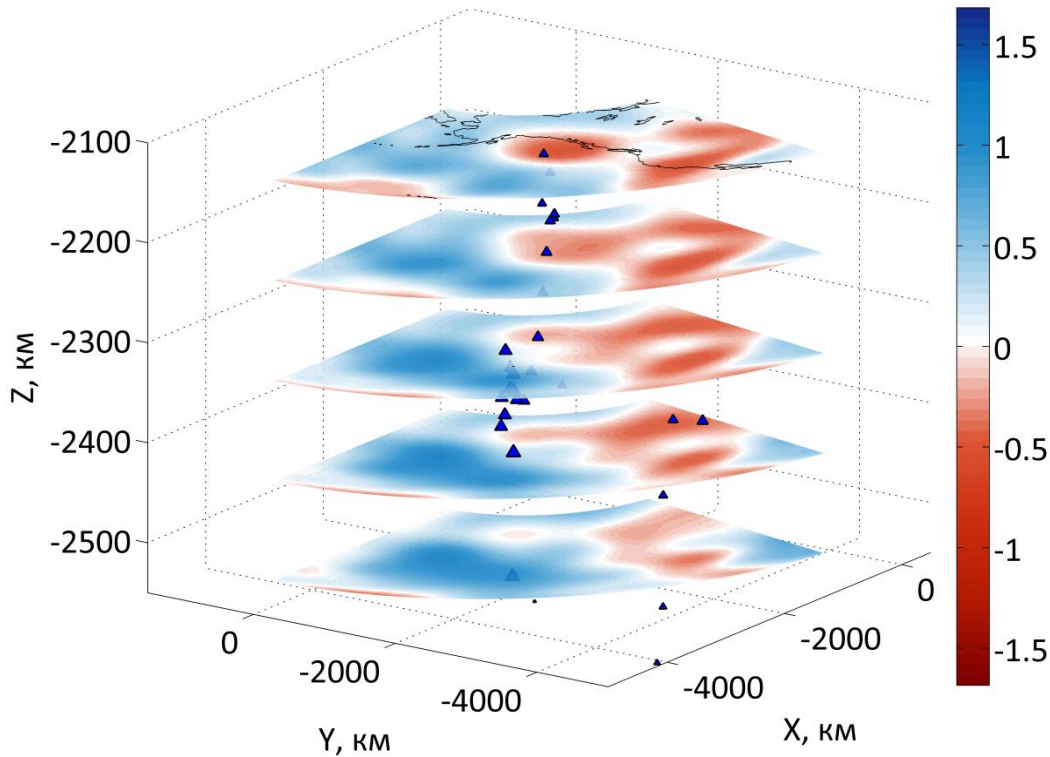

**Fig.7.** The lower-mantle structure calculated using the SMEAN2 model at the depth interval of 2100 to 2500 km. Legend is the same as for Fig. 3

## 4    Conclusions

The following main conclusions can be drawn:

- the substance of the liquid core, while mixing with the mantle substance, promotes the formation of hot channels;
- the kinematics of small-scale vortices which are suggested to be formed in the liquid core around the inhomogeneities of the core-mantle boundary can be used as an additional indicator of the presence of ascending channels in the lowest mantle;
- the remains of ancient lithospheric plates form the large inhomogeneities in the structure of the core-mantle boundary, which can significantly affect the structure of currents in the liquid core and lead to the formation and growth of the non-dipole part of the main geomagnetic field.